\documentclass[aps,pra,twocolumn,showpacs,amsmath,amssymb,floatfix]{revtex4}

\usepackage{graphicx}
\usepackage{dcolumn}
\usepackage{bm}
\usepackage[colorlinks,urlcolor=blue,citecolor=blue,linkcolor=blue]{hyperref}

\newcommand{\ket}[1]{\mbox{\ensuremath{\vert #1 \rangle}}}
\newcommand{\unit}[1]{\,\mathrm{#1}}
\newcommand{\partialD}[2]{\frac{\partial#1}{\partial#2}}
\newcommand{\Rb}{$^{87}$Rb\,}

\begin{document}

\title{Spatially inhomogeneous phase evolution of a two-component Bose-Einstein condensate}

\author{R. P. Anderson}
\author{C. Ticknor}
\author{A. I. Sidorov}
\author{B. V. Hall}
\email{brhall@swin.edu.au}
\affiliation{
ARC Centre of Excellence for Quantum-Atom Optics and \\
Centre for Atom Optics and Ultrafast Spectroscopy,\\
Swinburne University of Technology, Hawthorn, Victoria 3122, Australia
}

\date{\today}

\begin{abstract}
We investigate the spatially dependent relative phase evolution of an elongated two-component Bose-Einstein condensate. The pseudospin-$\tfrac{1}{2}$ system is comprised of the \ket{F=1,m_F=-1} and \ket{F=2,m_F=+1} hyperfine ground states of \Rb, which we magnetically trap and interrogate with radio-frequency and microwave fields. We probe the relative phase evolution with Ramsey interferometry and observe a temporal decay of the interferometric contrast well described by a mean-field formalism. Inhomogeneity of the collective relative phase dominates the loss of interferometric contrast, rather than decoherence or phase diffusion. We demonstrate a technique to simultaneously image each state, yielding subpercent variations in the measured relative number while preserving the spatial mode of each component. In addition, we propose a spatially sensitive interferometric technique to image the relative phase.
\end{abstract}


\maketitle

\section{Introduction}
Knowledge of the phase of matter waves is crucially important in studies of interferometry, entanglement, and precision measurement. Pertinent to all of these areas is the pseudospin-$\tfrac{1}{2}$ condensate, which can be realized with trapped neutral atoms in a superposition of two hyperfine ground states. Most commonly, the magnetically trappable \ket{F=1,m_F=-1} and \ket{F=2,m_F=1} states of \Rb have been used in experiments \cite{Hall98a,Hall98b,Matthews99,McGuirk03,Mertes07}. Early studies demonstrated spatial separation of the components \cite{Hall98a} and interferometric detection of relative phase for a region where both components remained overlapped following strongly damped center of mass motion \cite{Hall98b}. Also, the effect of phase winding throughout a two-component condensate was studied for continuous electromagnetic coupling \cite{Matthews99}. Following this, images of spin excitations in an ultracold uncondensed gas \cite{McGuirk02} enabled the study of spin domain growth for mixtures of condensed and uncondensed atoms \cite{McGuirk03}. The interference between two vortex lattices comprised of each component has also been examined \cite{Schweikhard04}. More recently, long-lived ringlike excitations of the binary condensate system have been observed \cite{Mertes07}. This kind of dynamical instability is accompanied by spatially dependent relative phase dynamics, which we investigate here. In particular, we consider the temporal decay of the interference signal obtained with a Ramsey-like measurement of the pseudospin-$\tfrac{1}{2}$ condensate.

The mechanism of phase diffusion for a two-component quantum gas has been recently studied \cite{Widera08}, whereby the evolution of a coherent spin state results in the decay of Ramsey visibility \cite{Sinatra00}. For the close inter- and intrastate interaction strengths in our system, phase diffusion is negligible \cite{Sinatra00}. Rather, we consider mean-field driven spatial inhomogeneities of the relative phase, which also act to decrease the interferometric contrast even without significant spatial separation between the components. This is relevant in the context of proposals to squeeze the macroscopic pseudospin in two-component quantum degenerate gases \cite{Sorensen01,Jin07,Rey07,Li08}, which are based on using the mean-field interaction as a source of entanglement, and development of a trapped atomic clock using an atom chip \cite{Treutlein04,Rosenbusch09}.

This paper is organized as follows. In Sec.~\ref{sec:experiment}, our experimental procedure is described up to the point of initialization of a two-component Bose-Einstein condensate (BEC). In  Sec.~\ref{sec:Ramsey_interferometry}, we report on the spatiotemporal relative phase dynamics of the two-component condensate using Ramsey interferometry. Simulations of coupled Gross Pitaevskii equations including atomic loss and electromagnetic driving terms yield striking agreement with observed matter wave interference. In Sec.~\ref{sec:dual_state}, we present the demonstration of a simultaneous state selective imaging technique. The method enables improved measurement of the longitudinal spin projection in a single experimental run, when classical fluctuations in relative and total atom numbers exist between different realizations of an experiment. Finally, in Sec.~\ref{sec:phase_interferometric}, we propose a combination of the dual state imaging method and Ramsey interferometry to directly image the spatial inhomogeneity and quantum fluctuations of relative phase.

\section{Realization of two-component BEC}
\label{sec:experiment}
A detailed description of our apparatus, including experiments performed with a \ket{F=2,m_F=2} condensate on a perpendicularly magnetized film atom chip, has been described elsewhere \cite{Hall06,Whitlock07,Hall07}. In brief, we use an atom chip with a machined Ag foil structure that allows currents to be passed in U- and Z-shape paths for surface magneto-optical trapping and Ioffe-Pritchard magnetic trapping \cite{Reichel99}. We begin by preparing a condensate in the \ket{F=1,m_F=-1} state, hereafter referred to as state \ket{1}. This is achieved in a manner similar to previous work with two main differences. First, we optically pump the atoms into \ket{1} prior to magnetic trapping using a $1\unit{ms}$ duration pulse of $\sigma^-$ polarized light, tuned to the D$_2$~($F =2 \rightarrow F' = 2$) transition, in the absence of repumping ($F = 1 \rightarrow F' = 2$) radiation.  To ensure the purity of \ket{1} during the magnetic trapping stage, a $2\unit{ms}$ pulse of optical pumping light illuminates the trapped cloud, completely removing residual magnetically trapped atoms in the $F = 2$ level. Second, the BEC is imaged after magnetic trapping via optical absorption using a $100\unit{\mu s}$ pulse of $\sigma^+$ light, tuned to the D$_2$~($F = 2 \rightarrow F' = 3$) transition. This is immediately preceded by a $1\unit{ms}$ pulse of repumping light that transfers all the atoms into the $F = 2$ manifold for imaging, while the short repumping-imaging delay ensures image blurring (from the recoil of a single repumping photon by each atom) is minimized. A charge-coupled device (CCD) camera records the absorption image using an achromat doublet lens, with a resolution of $7.5\unit{\mu m}$/pixel. Using this sequence, a pure \ket{1} BEC of $~ 2 \times 10^5$ atoms is routinely created in a cycle time of $40\unit{s}$.
\begin{figure}
    \includegraphics[width=1.0\columnwidth]{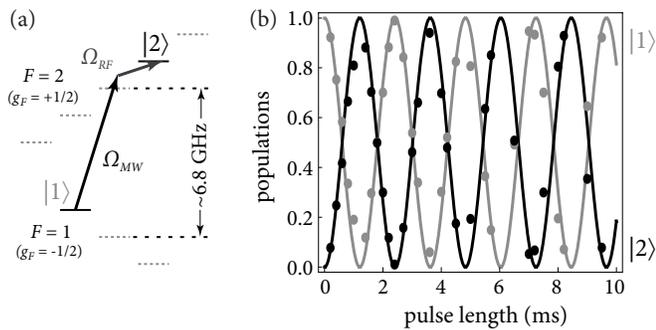}
    \caption{
    (a) Hyperfine ground states of \Rb. Two Zeeman levels are coupled by a two-photon microwave radio-frequency field (see text for description). (b) Two-photon Rabi oscillations of a pseduo-spin-$\tfrac{1}{2}$ BEC as measured by simultaneous detection of the populations of \ket{1} and \ket{2} (Sec.~\ref{sec:dual_state}).
    \label{fig:two_photon_Rabi}
    }
\end{figure}

Using a \ket{1} condensate, the radial and axial trap frequencies of the magnetic trap were measured to be \mbox{$f_{\rho} = 97.6(2)\unit{Hz}$} and \mbox{$f_z = 11.96(2)\unit{Hz}$}. We observe a \ket{1} condensate lifetime of $1.7\unit{s}$, well below the observed magnetic trap lifetime for a \ket{1} thermal cloud ($30\unit{s}$) and the expected three-body collisional-loss- dominated lifetime ($17\unit{s}$). Since the total atom number, BEC plus thermal cloud, is conserved, we attribute the observed lifetime to be limited by heating due to technical noise. The rate of this heating was measured to be $12(1)\unit{nK/s}$ which is small compared to the transition temperature of $120\unit{nK}$, yet ultimately of consequence. We prepare a two-component condensate of $1.5\times10^5$ atoms using a two-photon microwave radio-frequency field to couple $\ket{1}$ to $\ket{F = 2, m_F = 1}$, hereafter referred to as state $\ket{2}$. This is schematically presented in Fig.~\ref{fig:two_photon_Rabi}(a) and was initially demonstrated and described in Ref.~\onlinecite{Matthews98}. We use a magnetic trap with the potential minimum at $3.23\unit{G}$ (for which the relative Zeeman shift is independent of magnetic field to first order), as determined by microwave spectroscopy of a trapped condensate. The $\sim 6.8\unit{GHz}$ microwave radiation is derived from a signal generator (Agilent E8257D), pulsed using a fast ns switch (Agilent), then amplified (MA.Ltd $10\unit{W}$) and transmitted to the BEC using a helical antenna (gain $19\unit{dB}$) which resides in air $10\unit{cm}$ from the condensate. We simultaneously drive a radio-frequency magnetic field at $\sim 2.01\unit{MHz}$ with the atom chip wires used for evaporative cooling, albeit with an independent signal generator (SRS DS345) and an amplifier (MA.Ltd $2\unit{W}$). These fields couple \ket{1} and \ket{2} via the intermediate state \ket{F = 2, m_F = 0} with a detuning of $250\unit{kHz}$. This detuning ensures that no significant population is transferred to \ket{F = 2,m_F = 0}, a magnetically untrapped state, such that a two-level formalism adequately describes the multi-component dynamics of the system. We measure a two-photon Rabi frequency of $416\unit{Hz}$ [Fig.~\ref{fig:two_photon_Rabi}(b)] which is much faster than the characteristic time scale for the condensate to change shape. The high quality of these data is a result of the simultaneous state detection method described in Sec.~\ref{sec:dual_state}.

\section{Ramsey interferometry using two-component BEC}
\label{sec:Ramsey_interferometry}
Here we use Ramsey interferometry to study the relative phase evolution of a two-component condensate. The non-equilibrium dynamics observed in this system \cite{Mertes07} are accompanied by spatially dependent dynamics of the relative phase. We describe the evolution of the system using a pseudospinor formalism, whereby the internal and external states of the condensate are represented by a two-component order parameter
\begin{equation}\label{eq:spinor_definition}
\ket{\Psi ( \vec{r} ,t)} \equiv \left (
    \begin{array} {c}
    \Psi_1 ( \vec{r} ,t)\\
    \Psi_2 ( \vec{r} ,t)
    \end{array}
    \right)
    = \left (
    \begin{array} {c}
    \sqrt{n_{1}(\vec{r},t)} \, e^{i \phi_1 (\vec{r},t)}\\
    \sqrt{n_{2}(\vec{r},t)} \, e^{i \phi_2 (\vec{r},t)}
    \end{array}
    \right) \; ,
\end{equation}
where $n_1$ and $n_2$ are the atomic densities of each state in the condensate. We begin with a \ket{1} condensate in the ground state of the combined mean-field and external potentials, with density $n_0(\vec{r})$. This corresponds to the pseudospinor representation
\begin{equation}\label{eq:spinor_0}
    \ket{\Psi ( \vec{r} ,0)} =
    \left(
    \begin{array} {c}
    1\\
    0
    \end{array}
    \right)
    \sqrt{n_0(\vec{r})}
    \; ,
\end{equation}
Application of the first $\pi/2$ pulse (of length $t_{\pi/2}$) prepares the two-component superposition \cite{footnote1}
\begin{equation}\label{eq:spinor_initial}
    \ket{\Psi ( \vec{r} , t_{\pi/2})} =
    \frac{1}{\sqrt{2}} \left (
    \begin{array} {c}
    1\\
    -i
    \end{array}
    \right)
    \sqrt{n_0(\vec{r})}
    \; .
\end{equation}
This is no longer the ground state of the two-component system, as the mean-field interaction between component \ket{1} with component \ket{2}, and component \ket{2} with itself is different to that of component \ket{1} alone. This is due to a slight difference in the $s$-wave scattering lengths $a_{11} = 100.40\,a_0$, $a_{22} = 95.00\,a_0$, and $a_{12} = 97.66\,a_0$, where $a_0$ is the Bohr radius \cite{Mertes07}. Through the state inter-conversion, we modify the mean-field energy of the system by $\sim 1 \%$ for our experimental parameters and a condensate with $1.5\times10^5$ atoms. This is enough to drive hundreds of milliseconds of weakly damped collective excitations and coherent relative phase evolution. The system is allowed to evolve for a time $T$, after which we observe the condensate with or without the application of a second $\pi/2$ pulse
\begin{subequations}
\label{eq:spinor_final}
\begin{align}
    \ket{\Psi ( \vec{r} , t_{\pi/2} + T)} &=
    \left (
    \begin{array} {c}
    \sqrt{n_{1}(\vec{r})} \, e^{i \phi_1(\vec{r})}\\
    \sqrt{n_{2}(\vec{r})} \, e^{i \phi_2(\vec{r})}
    \end{array}
    \right)
    \; ; \label{eq:spinor_final_a} \\
    \ket{\Psi ( \vec{r} , t_{\pi/2} + T + t_{\pi/2})} &=
    \left (
    \begin{array} {c}
    \sqrt{n_{1}'(\vec{r})} \, e^{i \phi_1'(\vec{r})}\\
    \sqrt{n_{2}'(\vec{r})} \, e^{i \phi_2'(\vec{r})}
    \end{array}
    \right)
    \; . \label{eq:spinor_final_b}
\end{align}
\end{subequations}
where primes denote the densities and phases after the optional second pulse. We define a spatially dependent relative phase as
\begin{equation}\label{eq:phase_definition}
    \phi(\vec{r}) \equiv \phi_2 (\vec{r}) - \phi_1 (\vec{r}) \; .
\end{equation}
In the frame rotating at the effective frequency of the two-photon coupling, this phase evolves to
\begin{equation}\label{eq:phase_evolved}
    \phi(\vec{r}) = \Delta \, T + \phi_{\text{mf}}(\vec{r}) + \delta \phi
    \; ,
\end{equation}
after evolution time $T$, where $\Delta$ is the detuning of the two-photon field from the atomic resonance, $\phi_{\text{mf}}$ is the spatially dependent phase whose evolution is driven by the mean field, and $\delta \phi$ is any phase shift intentionally applied to the coupling field before $t = t_{\pi/2} + T$. Varying $\Delta$, $T$, or $\delta \phi$ and applying a second $\pi/2$ pulse yields a Ramsey interference signal $\alpha$, which we define as the expectation value of the longitudinal psuedo-spin projection, scaled to vary as the normalized population difference
\begin{equation}\label{Sz_definition}
    \alpha \equiv (N \hbar)^{-1} \langle \hat{S_z} \rangle
    = \left( N_2' - N_1' \right)/N \; ,
\end{equation}
where $N_i' = \int n_i' \, d \vec{r} \, , (i = 1,2)$ and $N$ is the total atom number.

\begin{figure}
    \includegraphics[width=1.0\columnwidth]{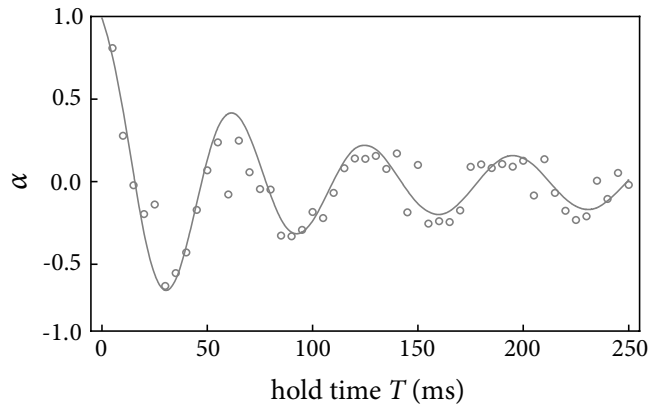}
    \caption{The Ramsey interference signal of a pseudospin $^{87}$Rb BEC where the detuning of the two-photon field from the atomic resonance $\Delta/2\pi = 1 \unit{Hz}$. The points are experimental measurements and the solid line is calculated using the CGPE theory, with an additional exponential decay due to decoherence \cite{footnote2}. The $\sim 16\unit{Hz}$ oscillation, decay, and frequency chirp of $\alpha$ are predominantly due to spatial evolution of the relative phase.
    \label{fig:Ramsey_data_cf_GP_2.5Hz}}
\end{figure}

A typical Ramsey signal obtained by varying the evolution time $T$ is shown in Fig.~\ref{fig:Ramsey_data_cf_GP_2.5Hz}. We observe nonexponential decay and chirp to Ramsey fringes in the time domain. These observations are consistent with the spatial evolution of a coherent two-component order parameter, as described below. Decoherence does not significantly contribute to the decay of Ramsey contrast we observe \cite{footnote2}. It is the spatial dependence of the mean-field driven phase that leads to a decrease in contrast of Ramsey fringes over time. It should be noted that this dephasing is unlike the inhomogeneous dephasing that occurs in statistical ensembles of uncondensed thermal atomic samples. Rather, it is the variation of the relative phase across a many-body wave function.

To simulate the evolution of density and relative phase, the dynamics throughout electromagnetic coupling pulses, and the Ramsey interferometry signal, we solve the coupled Gross-Pitaevskii equations (CGPE) for both components with decay terms corresponding inter- and intrastate many-body loss processes. In their most basic form (without loss or electromagnetic coupling terms), these equations were introduced in \cite{Ho96,Zeng95}. The inclusion of electromagnetic coupling terms has appeared in \cite{Ballagh97} (multicomponent condensates) and \cite{Marzlin97,Dum98,Williams99,Blakie99,Merhasin05,Gordon08} (specifically pseudospin-$\tfrac{1}{2}$ condensates). Addition of nonlinear terms to describe many-body loss processes was used in \cite{Yurovsky99} and applied to a binary condensate system of \Rb in \cite{Mertes07}. Combining all of the above features, we arrive at
\begin{subequations}
\label{eq:CGPEs}
\begin{align}\label{eq:CGPE1}
    \lefteqn{ i \hbar \partialD{\Psi_1}{t} = \frac{\hbar \, \Omega}{2}\Psi_2+}
    \\ & &
    \left[ -\frac{\hbar^2 \nabla^2}{2m} +
    V_1 + g_{11} |\Psi_1|^2 + g_{12} |\Psi_2|^2
    - i \hbar \, \Gamma_1 \right] \Psi_1
    \, , \notag
\end{align}
and
\begin{align}\label{eq:CGPE2}
    \lefteqn{ i \hbar \partialD{\Psi_2}{t} = \frac{\hbar \, \Omega}{2}\Psi_1 - \hbar \, \Delta \Psi_2 +}
    \\ & &
    \left[ -\frac{\hbar^2 \nabla^2}{2m} +
    V_2 + g_{22} |\Psi_2|^2 + g_{12} |\Psi_1|^2
    - i \hbar \, \Gamma_2 \right] \Psi_2
    \, . \notag
\end{align}
\end{subequations}
where $m$ is the mass of \Rb, $V_i$ is the magnetic trapping potential experienced by component $i$, and \mbox{$g_{ij} = 4\pi \hbar^2 a_{ij}/m$} are the mean-field interaction parameters. The dominant two- and three-body loss processes are described by the terms \mbox{$\Gamma_1=(\gamma_{111}n_1^2 + \gamma_{12} n_2)/2$} and  \mbox{$\Gamma_2=(\gamma_{22} n_2 +\gamma_{12} n_1)/2$}, with loss rates $\gamma_{i..j}$ measured and described in Ref.~\onlinecite{Mertes07}.

To solve Eqs.~(\ref{eq:CGPEs}) numerically, we exploit the cylindrical symmetry of our experimental geometry, and assume that the wave function takes the form \mbox{$\Psi(\vec{r}) \rightarrow e^{i \, m \, \varphi} \Psi_m(\rho , z)$,} where $m$ is an integer. Since the ground state of the system has $m = 0$, we exclusively solve for this case and assume the time evolution does not break this symmetry. We use the discrete Hankel-Fourier transform to solve the CGPE and have adapted the procedure used in \cite{Ronen06} for multicomponent systems. Essentially, the technique facilitates the solution of second-order partial differential equations in cylindrical coordinates using spectral methods, where the Fourier transform  fails due to divergence of the Laplacian for zero radial momentum. Implementation of the discrete Hankel transform is greatly simplified by sampling the fields at zeros of the $m^{\text{th}}$ order Bessel function of the first kind, rather than using a Cartesian grid.
\begin{figure}
    \includegraphics[width=0.9\columnwidth]{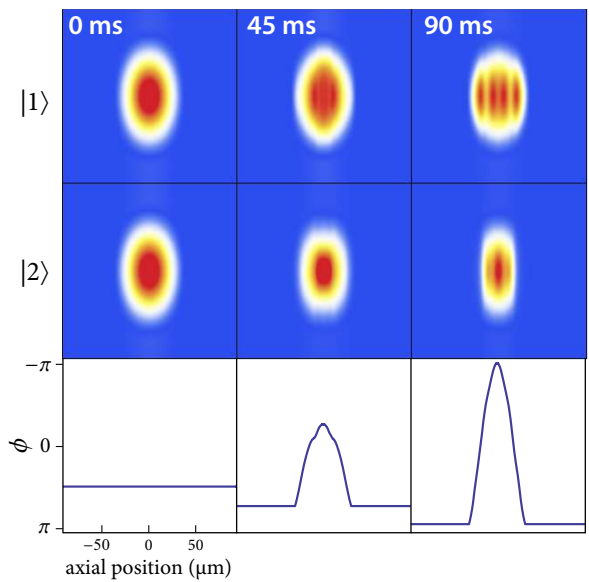}
    \caption{Simulated in-trap column densities $\int n_i \, dx$ and relative phase $\phi(\rho=0,z)$ [see Eqs.~(\ref{eq:spinor_final}) and (\ref{eq:phase_definition})] for various evolution times of the initial state in Eq.~ (\ref{eq:spinor_initial}). The initially uniform phase exhibits inhomogeneous evolution, accompanied by density excitations in the axial direction $(z)$. An offset of $14\, \pi\unit{rad\,s^{-1}}$ has been added to the relative phase for clarity.
    \label{fig:phase_evolution}}
\end{figure}

In Fig.~\ref{fig:phase_evolution}, we show the simulated column densities of each component for various evolution times following preparation of the initial superposition in Eq.~(\ref{eq:spinor_0}). The relative phase is seen to vary across the axial dimension of the condensate. This relative phase modulation eventually becomes manifest as relative density modulation, such that the components spatially separate along the axial direction, consistent with the sign of $a_{11} a_{22} - a_{12}^2$ that governs the miscibility of two-component condensates \cite{Ho96}. This is evident in Fig.~\ref{fig:phase_evolution} at $90\unit{ms}$, where \ket{1} predominantly resides further from the origin than \ket{2}. In the absence of loss, the moments $\int \vert z \vert n_i \, d\vec{r}$ would continue to oscillate out of phase, with \ket{2} occupying a larger volume than \ket{1} at $\sim 200 \unit{ms}$. However, atomic loss dampens the collective oscillations on this time scale.

\begin{figure*}
    \includegraphics[width=2.05\columnwidth]{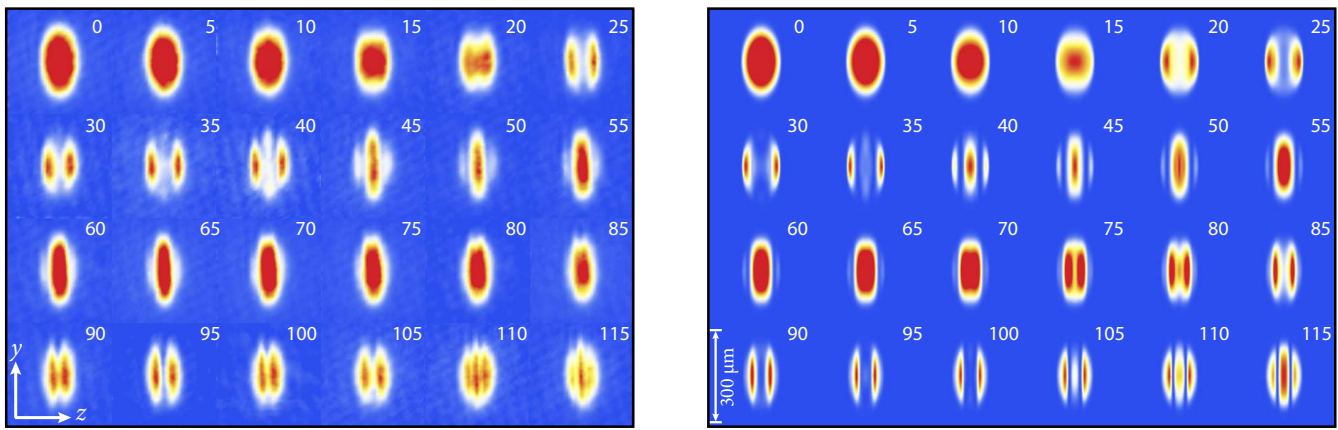}
    \caption{Column density of state \ket{2} as a function of Ramsey interferometry time (inset for each image in milliseconds). Application of the second $\pi/2$ pulse locally converts the relative phase into relative population. (Left) Experimental: single shot absorption images, taken after a $20\unit{ms}$ fall time, reveal the mean-field driven phase evolution. (Right) Theoretical: column density plots obtained by solving the coupled Gross-Pitaevskii equations show excellent agreement with experimental results.
    \label{fig:Exp_Vs_GP}}
\end{figure*}

We present here measurements of the density of state \ket{2} for various evolution times $T$, followed by a second $\pi/2$ pulse. We observe striking agreement between these results and those of CGPE simulations. In Fig.~\ref{fig:Exp_Vs_GP}, the column density $\int n_2 \, dx$ is shown, where $x$ is the imaging axis, a component of the radial coordinate (of tight confinement) $\rho$. The only free parameter used in the simulation is the detuning $\Delta$ of the two-photon field from the atomic transition. We adjust this parameter to optimize the agreement (in this case $\Delta/2\pi = 1\unit{Hz}$), consistent with a measurement of the detuning using a cold thermal cloud. In particular, the three distinct density maxima at $40$ and $115\unit{ms}$, corresponding to antinodes of the state \ket{2} wave function, are represented accurately by the simulation. This prominent density modulation along the $z$ axis (of weak confinement) reveals the nonuniform relative phase variation along this direction. This is enabled by the second $\pi/2$ pulse, which converts information about the relative phase into (relative) density. In the absence of the second pulse, while density modulation still exists, it is of lower contrast (Fig.~\ref{fig:phase_evolution}, $90\unit{ms}$ is the most structure we observe from simulations).

Furthermore, the Ramsey contrast decays significantly faster than the components spatially separate. We emphasize that uncommon spatial modes of the components are neither necessary nor sufficient for loss of interferometric visibility. For example, consider the ground state of the two-component system, whose density consists of a central feature of state \ket{2}, enclosed by two lobes of state \ket{1}. Of course, no inhomogeneous relative phase evolution occurs for the ground state and the mean-field limited Ramsey visibility is $80 \%$. This remains true even in the presence of loss, which we have confirmed from simulations that show the condensate wave functions adiabatically follow the time-dependent ground state.

We attribute residual differences in the precise spatial structure of the measured and simulated macroscopic wave functions to the finite resolution of our imaging system, photon recoil blurring during the imaging pulse, imperfect cylindrical symmetry of the trap, and interaction of the condensate with an undetectable yet significant thermal component for longer times. The repeatability of the observed interference fringes (both run-to-run \textit{and} day-to-day) is indicative of the well-defined initial-state preparation and phase evolution in this system.

\section{Dual State Imaging}
\label{sec:dual_state}
We present a new imaging technique to simultaneously detect states $\ket{1}$ and $\ket{2}$ with spatial resolution. This represents the first unambiguous state selective measurement of a \mbox{pseudospin-$\tfrac{1}{2}$} condensate that preserves the spatial mode. pseudospin-$\tfrac{1}{2}$ condensates of \Rb suitable for interferometry are commonly comprised of states whose magnetic moments are similar, complicating spatial separation of the constituent states with a magnetic field gradient. Various methods exist to individually image either the $F=1$ or $F=2$ populations, using resonant light in combination with optical or microwave repumping pulses. An alternative method uses a radio-frequency sweep to distribute the $\ket{1}$ and $\ket{2}$ populations among all eight Zeeman sublevels which are then spatially resolved using Stern-Gerlach separation prior to imaging. The original \ket{1} and \ket{2} populations are inferred from estimates of how the adiabatic rapid passage distributes population among the sublevels \cite{Streed06}. Phase-contrast imaging can also been used to image the difference in populations \cite{Matthews99} and has enabled the observation of vortex dynamics \cite{Anderson00,Anderson01,Schweikhard04}.
\begin{figure}
    \includegraphics[width=0.75\columnwidth]{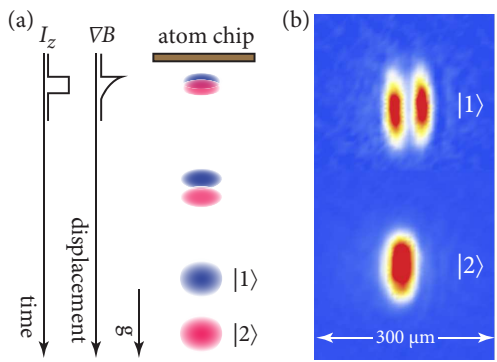}
    \caption{
    Transfer of \ket{1} to the magnetic field insensitive state \ket{2,0} is achieved by adiabatically sweeping the atomic resonance in the presence of a fixed frequency microwave pulse. (a) Spatial separation is performed by pulsing $I_z$ (see text for description) to induce a Stern-Gerlach force that accelerates \ket{2} away from the atom chip and \ket{1}. (b)~Absorption imaging using the D$_2$ ($F = 2 \rightarrow F' = 3$) transition yields a single shot image of a two-component condensate, \ket{1} (above) and \ket{2} (below), where the \ket{1} structure arises during a Ramsey interference experiment described in Sec.~\ref{sec:Ramsey_interferometry}.
    \label{fig:dual_scheme}}
\end{figure}

Our dual state detection method consists of three stages. First, the magnetic trap is switched off by reducing the Z-wire current, $I_z$, to zero in $\sim 200 \unit{\mu s}$. The uniform bias field that completes the trapping potential is concurrently ramped off in $3\unit{ms}$ and a fixed frequency microwave field is applied. The Zeeman splitting of the magnetic sublevels decreases, facilitating rapid passage from \ket{1} to \ket{F = 2, m_F = 0}. This population transfer method is more robust than a resonant microwave $\pi$ pulse, owing to the less stringent requirements of rapid passage on the stability of magnetic fields and microwave power. The same microwave field is used for the two-photon pulses, obviating the need for a second microwave frequency synthesizer and enabling detection immediately after a two-photon pulse, as in Sec.~\ref{sec:Ramsey_interferometry}. For our initial experiments demonstrating this technique, we transferred $44(1)\%$ of \ket{1} population to \ket{F = 2, m_F = 0} (rapid bias field switching was desired at the expense of adiabaticity). At this point, both spin components of the condensate are spatially overlapped and fall freely under gravity. However those in the \ket{F = 2, m_F = 0} state are largely insensitive to magnetic fields.

The second step [Fig.~\ref{fig:dual_scheme}(a)] produces spatial separation
of the two components. This is achieved using a $30\unit{A}$, $1\unit{ms}$ $I_z$ pulse which induces a Stern-Gerlach force in the direction of gravity to displace \ket{2} such that both components are separated by $\sim~300\unit{\mu m}$ after $20\unit{ms}$ of free fall. Finally, an absorption image is taken
[Fig.~\ref{fig:dual_scheme}(b)] yielding the column density of \ket{2} directly below \ket{1}. This image is from a Ramsey interferometry experiment described in Sec.~\ref{sec:Ramsey_interferometry} and highlights the ability of this dual state imaging method to preserve spatial information about the wave functions. This is due to the minuscule recoil velocity ($5 \times 10^{-9} \unit{mm/s}$) associated with microwave photon absorption thus eliminating the image blur associated with optical repumping techniques.

\begin{figure}
    \includegraphics[width=1.0\columnwidth]{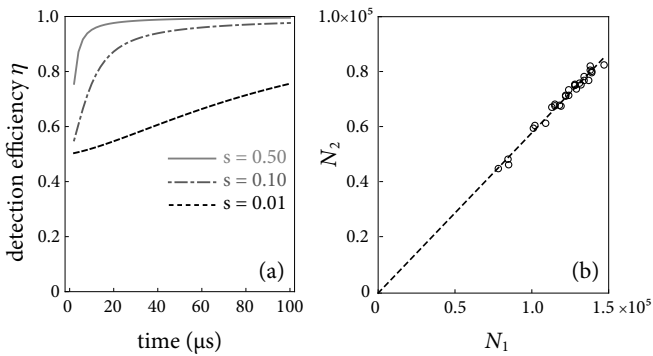}
    \caption{(a) The detection efficiency $\eta$ for dual state absorption imaging as a function of imaging pulse duration, for three typical saturation parameters, $s=0.50$ (solid), $s=0.10$ (dash-dotted) and $s=0.01$ (dashed). (b) Simultaneous measurement of atoms in \ket{1} and \ket{2} after a two-photon coupling pulse showing high correlations for a range of detected BEC atom numbers. An imaging pulse of $100\unit{\mu s}$ duration with $s = 0.1$ was used.
    \label{fig:detection_efficiency}}
\end{figure}

While conservation of spatial information is crucial for dynamical studies of two-component BEC, the detection method should also determine the atom number accurately. We investigate the effects of optical pumping during absorption imaging on $\eta$, the relative efficiency of detecting atoms originating in \ket{F = 2, m_F = 0} to those starting in \ket{2}. The absorption of a $\sigma^+$ polarized probe beam was modeled using the optical Bloch equations \cite{Maguire06}, including the Zeeman splitting of the ground and excited states. We evaluated the absorption by integrating the total excited-state population ($F' = 1,2,3 \, ; m_F' = - F',..,F'$) during the imaging pulse for a given initial state. Results are shown in Fig.~\ref{fig:detection_efficiency}(a) for different saturation parameters $s = I_0 / 1.68 \unit{mW\,cm^{-2}}$ of the $\ket{F = 2, m_F = 2} \rightarrow \ket{F' = 3, m_F' = 3}$ cycling transition, with $I_0$ the intensity of the probe beam. As the imaging pulse duration increases, optical pumping acts to populate the \ket{F=2,m_F=2} state and more time is spent driving the cycling transition. As such, the relative detection efficiency asymptotes to unity and does so faster for higher imaging intensities. For our typical imaging conditions, we predict a relative detection efficiency of $\eta = 0.98$.

The dual state imaging method minimizes fluctuations in measured relative number that arise when detecting individual populations in different experimental realizations. This is due to common mode rejection of both probe-laser frequency noise and irreproducibility of total atom number between experimental runs. This is demonstrated in Fig.~\ref{fig:detection_efficiency}(b) which shows a strong correlation between \ket{1} and \ket{2} populations for a wide range of detected total condensate atom number. The method allows the relative population to be measured with a subpercent standard deviation, as opposed to single-component measurements of the absolute population which can typically vary by $\pm 10 \%$. This method improves our measurement of phenomena such as Rabi oscillations [Fig.~\ref{fig:two_photon_Rabi}(b)] and Ramsey fringes (Fig.~\ref{fig:Ramsey_data_cf_GP_2.5Hz}). Moreover, it will see important application in atom chip clock research, allowing normalization of total atom number fluctuations, a practice used to enhance the accuracy of precision fountain atomic clocks \cite{Rosenbusch09,Santarelli99}. It also represents a significant step to achieving quantum-limited detection of the total spin projection in pseudospin-$\tfrac{1}{2}$ condensates.

\section{Phase reconstruction using Ramsey interferometry}
\label{sec:phase_interferometric}
\begin{figure}
    \includegraphics[bb=16 0 325 306,width=0.9\columnwidth,clip=true]
    {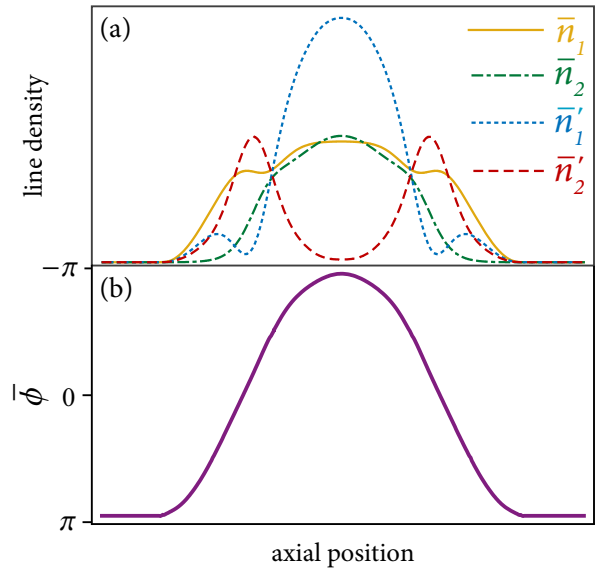}
    \caption{Simulation of interferometric phase retrieval
    method. (a) Two absorption images are captured for each evolution time,
    yielding the line densities immediately prior to, and after the second $\pi/2$ pulse: $\bar{n}_1$, $\bar{n}_2$, and $\bar{n}'_1$, $\bar{n}'_2$
    respectively.  (b) Point-by-point arithmetic is used [Eq.~(\ref{eq:phase_interferometric_LD})] to extract the relative phase profile along the axial direction.
\label{fig:phase_reconstruct}}
\end{figure}

In this system, imaging the relative phase amounts to resolving the transverse component of the local Bloch-vector corresponding to the pseudospinor in Eq.~ (\ref{eq:spinor_definition}). For a sample of ultracold uncondensed \Rb atoms, this has been achieved in Ref.~\onlinecite{McGuirk02}. There, the Ramsey interference signal was binned in both space and time, recovering a phase spectrogram that illustrated collective spin excitations driven by exchange scattering. This requires the repetition of many experimental cycles and gives an averaged relative phase with reduced temporal resolution. We propose a technique to spatially resolve the relative phase of the pseudospin-$\tfrac{1}{2}$ condensate from two dual state absorption images and present theoretical results that investigate the method's applicability and limitations. Performing an identical experiment with and without the presence of the second $\pi/2$ pulse permits four column densities $\tilde{n} = \int n \, dx$ to be imaged, where $n \in \left\{ n_1, n_2, n'_1, n'_2 \right\}$ and the densities are defined in Eqs.~(\ref{eq:spinor_final}). The densities can be used to recover the relative phase $\phi(\vec{r})$ using
\begin{equation}\label{eq:phase_interferometric}
n'_2(r)-n'_1(r)= 2\sqrt{n_1(r)n_2(r)} \sin(\phi(\vec{r})) \, .
\end{equation}
There is little variation of the relative phase along the radial direction, and the densities of both components have similar radial dependence. This is a consequence of the tight radial confinement, as the energy of the system is not sufficient for significant radial excitations to be induced. As such, the radially integrated densities, the \textit{line densities} $\bar{n} = \int n \, 2\pi \rho \, d\rho = \int \tilde{n} \, dy$, are representative of the relative phase evolution, such that
\begin{equation}\label{eq:phase_interferometric_LD}
\bar{n}'_2(z)-\bar{n}'_1(z) = 2\sqrt{\bar{n}_1(z)\bar{n}_2(z)} \sin(\bar{\phi}(z))
\end{equation}
defines a phase $\bar{\phi}(z)$ that is a good estimate of the phase along the center of the condensate $\phi(\rho = 0, z)$. We have confirmed this by evaluating Eq.~(\ref{eq:phase_interferometric_LD}) with line densities attained from simulations, as shown in Fig.~\ref{fig:phase_reconstruct}. For $90\unit{ms}$ of evolution, where the range of relative phase is $1.9 \, \pi$ across the condensate, the agreement is within $78\unit{mrad}$.
In more elongated geometries or where tunable miscibility can be realized, radial excitations are further suppressed and the approximation improves. In these cases there is increased spin locking along the imaging axis and the integration of density inherent to absorption imaging corrupts the phase signal less.

The requirements of this technique are more demanding than those of the experimental investigation presented in Sec.~\ref{sec:Ramsey_interferometry}. For successful implementation of Eq.~\ref{eq:phase_interferometric_LD} to point-by-point arithmetic on multiple absorption images, we are currently pursuing improved detection methods with higher signal-to-noise ratio, spatial resolution, and dynamic range. This technique may be useful for investigations of spin squeezing, in the context of recent proposals which suggest using the nonlinear interaction to reduce the spin projection noise below the standard quantum limit \cite{Sorensen01,Jin07,Rey07,Li08}. It is interesting to consider the spatial dependence of the relative phase \textit{fluctuations} of a macroscopic pseudospinor. This technique may allow such quantum noise to be studied with spatial resolution.

\section{\label{sec:conclusion}Conclusion}
We have quantitatively studied the spatial evolution of the relative phase of a pseudospin-$\tfrac{1}{2}$ $^{87}$Rb BEC. Preparing the system with a $\pi/2$ pulse yields a non-equilibrium state due to small differences in the inter- and intra-state scattering lengths $a_{11}$, $a_{22}$, and $a_{12}$. We demonstrate the resulting spatially inhomogeneous evolution of the collective relative phase. This acts to decrease the Ramsey contrast before significant spatial separation of the components. Spatial dephasing of the many-body wave function dominates the effects of decoherence and phase diffusion on the interferometric contrast. A second $\pi/2$ pulse maps the non-uniform phase distribution onto the population distribution of each spin state. Measuring the longitudinal pseudospin projection is enhanced by a new dual state imaging method that combines microwave adiabatic rapid passage and Stern-Gerlach spin filtering. This yields a single shot measurement of \ket{1} \textit{and} \ket{2}, enabling the relative populations to be measured with subpercent precision. Moreover, this novel method does not corrupt the spatial modes of either component. We propose extending this technique to image the relative phase profile of an elongated \mbox{pseudospin-$\tfrac{1}{2}$} BEC to study the spatial dependence of relative phase fluctuations.

\begin{acknowledgments}
We wish to thank Chris Vale, Peter Hannaford, Simon Haine, Mattias Johnsson, and \mbox{Peter Drummond} for useful discussions. This project is supported by the ARC Centre of Excellence for Quantum-Atom Optics, and an ARC LIEF grant LE0668398.
\end{acknowledgments}

\bibliography{rpa_manuscript}

\begin{thebibliography}{38}
\bibitem[{foo({\natexlab{a}})}]{footnote1}
\bibinfo{note}{This assumes that there is no evolution of the spatial modes
  during the coupling pulse, i.e. the pulses perform spin rotations locally
  throughout the condensate. This is valid for sufficiently strong
  electromagnetic coupling of the states, namely, when $\Omega \gg
  t_{\text{heal}}^{-1}$, where $t_{\text{heal}}$ is the characteristic time for
  the condensate spatial mode to respond to the mean-field perturbation. We
  have confirmed the validity of this approximation by simulating the
  condensate evolution during the coupling pulse using the CGPE, Eqs.
  (\ref{eq:CGPEs}). Typical parameters lead to a phase variation of less than
  $15\unit{mrad}$ across the condensate. This is sufficient to consider the
  relative phase spatially uniform across the condensate immediately following
  the first $\pi/2$ pulse.}
\bibitem[{foo({\natexlab{b}})}]{footnote2}
\bibinfo{note}{We infer a decoherence time of $350\unit{ms}$, which is the
  exponential decay required to account for the loss of Ramsey visibility in
  Fig.~\ref{fig:Ramsey_data_cf_GP_2.5Hz}, in addition to the dominant
  mean-field dephasing.}
\end{thebibliography}

\end{document}